# Synthesis of built-in highly strained monolayer MoS$_2$ using liquid precursor chemical vapor deposition

*L. Seravalli[1], F. Esposito[1,2], M. Bosi[1], L. Aversa[3], G. Trevisi[1], R. Verrucchi[3], L. Lazzarini[1], F. Rossi[1], F. Fabbri[4]*

[1] Institute of Materials for Electronics and Magnetism (IMEM-CNR), Parco Area delle Scienze 37/a, 43124 Parma, Italy

[2] Department of Mathematical, Physical and Computer Sciences, University of Parma, Parco Area delle Scienze 7/a, 43124 Parma, Italy

[3] Institute of Materials for Electronics and Magnetism (IMEM-CNR), FBK Trento unit, Via alla Cascata 56/C, 38123 Povo (Trento), Italy

[4] NEST, Istituto Nanoscienze – CNR, Scuola Normale Superiore, Piazza San Silvestro 12, 56127 Pisa, Italy

**Abstract**

Strain engineering is an efficient tool to tune and tailor the electrical and optical properties of 2D materials. The built-in strain can be tuned during the synthesis process of a two dimensional semiconductor, as molybdenum disulfide, by employing different growth substrate with peculiar thermal properties. In this work we demonstrate that the built-in strain of MoS$_2$ monolayers, grown on SiO$_2$/Si substrate using liquid precursors chemical vapor deposition, is mainly dominated by the size of the monolayer. In fact, we identify a critical size equal to 20 μm, from which the built-in strain increases drastically. The built-in strain is maximized for 60 μm sized monolayer, leading to 1.2% tensile strain with a partial release of strain close to the monolayer triangular vertexes due to formation of nanocracks. These findings also imply that the standard method for evaluation of the number of layers based on the Raman modes separation becomes unreliable for monolayer with a lateral size above 20 μm.

**Introduction**

Strain engineering is an efficient tool to tune and tailor the electrical and optical properties of 2D materials since their electronic band structures are highly sensitive to mechanical deformation.[1–8] Among the different classes of 2D materials, semiconducting transition metal dichalcogenides (TMDs) have demonstrated the most interesting and surprising modifications of electronic properties induced by the application of strain.[9,10] For instance, strain can induce the indirect-to-direct bandgap transition in multilayer WSe$_2$ flake[11], while the opposite transition (the direct-to-indirect transition) can occur in monolayer TMDs[12,13]. With regard to the optical properties the local strain application can either induce exciton funneling[14,15], efficient exciton to trion conversion[16] or the formation a new hybrid state of dark and localized excitons[17].

It is worth noting that TMD monolayers can withstand high tensile strains before breaking, as high as 10% for molybdenum based[18,19] and 19% for tungsten based[20] exfoliated monolayers.

Different approaches have been employed for the application of external stress to 2D materials: the most employed one is based on the employment of bendable and stretchable polymeric substrates.[21–27], where it is possible to apply mainly uniaxial strain to the 2D materials. This approach is maximized in the fabrication of the origami-like[28] and kirigami-like[29] MoS$_2$ based devices, such as strain sensors or optoelectronic devices. A similar approach has been employed to fabricate human eye mimicking photodetectors.[30] Another widely used method is the transfer of the 2D material on a patterned non-planar substrate for inducing localized strain, resulting into a local change of the band structure. The substrate can present insulating[31–34] or semiconducting[35,36] structures, namely stressors, that can apply a large local strain degree at their top. This large localized strain application modifies the band structure of 2D semiconducting materials, enabling efficient charge collection, desirable for bright single photon emission.[33,34] High degree of strain is also achieved by suspending 2D membranes on hole patterned substrates.[37] A local strain increase can be achieved on suspended membranes using different approaches: using an AFM tip[38], applying an external gas pressure[39] or applying a gate voltage between a suspended monolayer membrane and an electrode below[17]. An additional novel approach to apply strain to TMDs heterostructures is the employment of polymeric artificial muscles, that taking advantage of low friction between different 2D materials can apply a tensile strain in van der Waals heterostructures.[40]

The strain tuning during the chemical vapor deposition (CVD) process of 2D materials has been also demonstrated.[41,42] This method relies on the mismatch of thermal expansion coefficient (TEC) mismatch between the substrate and the 2D material.[4] For instance, monolayer WSe$_2$ with built-in strains ranging from 1% tensile to compressive 0.2% was obtained using different growth substrates. The TEC approach has been also employed in order to synthetize strained WS$_2$ on quartz with an oriented array of wrinkles.[43] In addition, strain induced buckling has been recently identified as a possible cause of threefold symmetric domain formation in hexagonal shaped WS$_2$ monolayers.[44] In case of MoS$_2$, the modification of strain and doping using different growth substrates has been demonstrated, where the maximum built-in tensile strain is found to be 0.4% with a SiO$_2$ growth substrate.[45]

In this work we demonstrate the dependence of the tensile strain as function of the lateral size of sharp-vertex shaped MoS$_2$ monolayers grown on standard 300 nm thick SiO$_2$/Si substrate using liquid precursor chemical vapor deposition. The built-in strain is demonstrated and evaluated by scanning Raman and photoluminescence spectroscopy. The monolayer MoS$_2$, with lateral size below 20 μm, presents a built-in compressive strain of 0.3%, while the flakes with the largest lateral size (60 μm) are more strained, reaching an upper limit of 1.2%. It is worth noting that such flakes are affected by nanocracks close to the vertexes, revealing a partial release of the strain down to 0.7%. The built-strain affects the optical properties; in particular, the highly strained flakes show a quenching and a red-shifting of the excitonic emissions of MoS$_2$.

**Results and discussion**

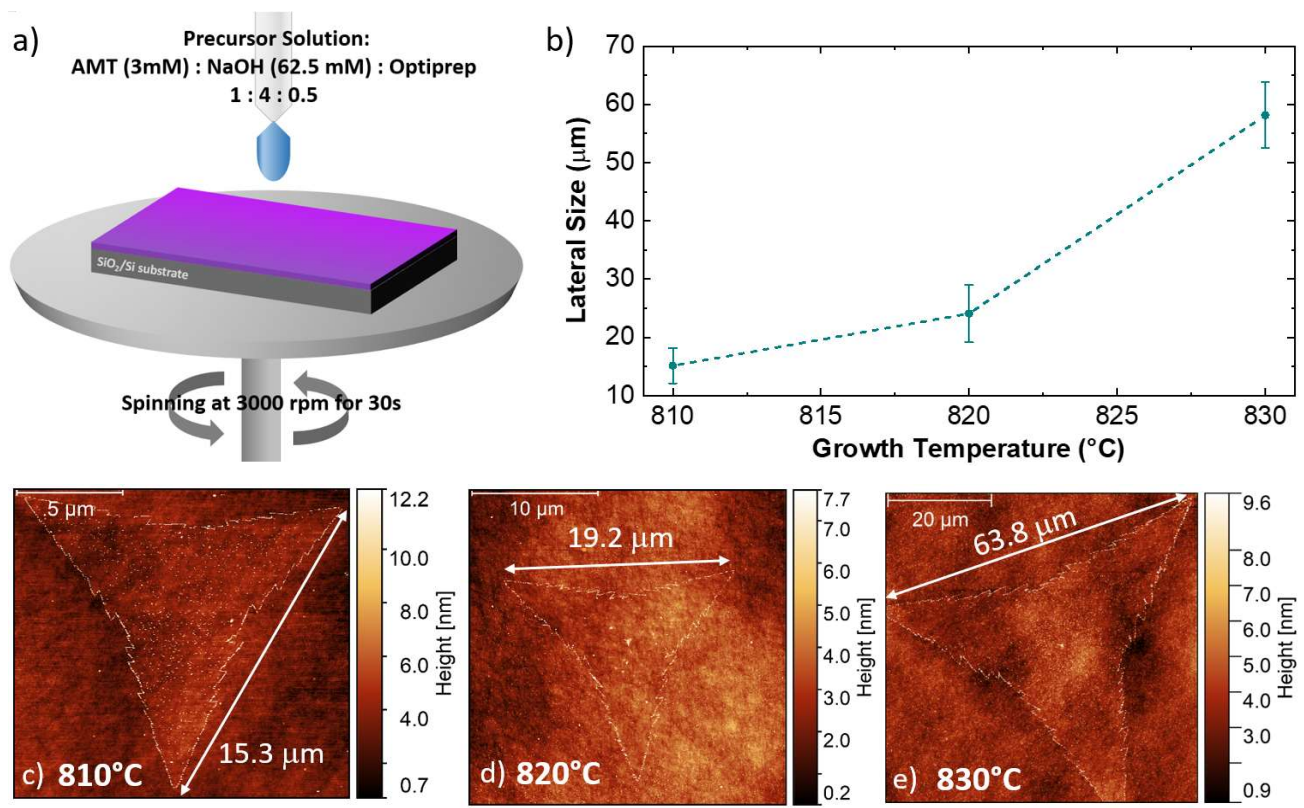

**Figure 1. a)** Illustrative sketch of the substrate preparation prior to the growth process. **b)** Resuming graph of the MoS$_2$ monolayer lateral size versus the growth temperature of the CVD process. **c), d) and e)** Representative AFM topographic maps of the MoS$_2$ flake grown at different temperatures. The lateral size of the flake (white arrow) is 15.3 μm, 19.2 μm and 63.8 μm in case of the 810°C, 820°C and 830 °C growth temperatures, respectively.

Figure 1 illustrates the outcome of the CVD synthesis at increasing temperature using a solution containing molybdenum liquid precursor. In particular, the sketch (Fig 1a) shows the deposition of the precursor solution on the SiO$_2$/Si growth substrate by spin coating, carried out for 30 s at 3000 rpm, of the precursor solution. This solution has three main components: a 3 mM solution of ammonium molybdate tetrahydrate (AMT), a 62.5 mM solution of sodium hydroxide (NaOH) and a 0.4 mM solution of iodixanol (Optiprep). The atomic force microscopy (AFM) is employed to study the number of layers composing the flakes and their lateral size. It is worth noting that the flakes obtained with the liquid precursor CVD have the sharp-vertex triangular shape.[46] The lateral size data are reported in Fig. 1b, that highlights the increase of the flake size with the growth temperature. The flakes, grown at 810°C, present an average lateral size of (15.1 ± 3.0) μm, while the flakes synthetized at 820°C present an average lateral size of (24.1 ± 4.9) μm. The maximum lateral size, (58.2 ± 5.7) μm is obtained when the growth process is carried out at 830°C. Fig 1c, 1d and 1e present representative AFM topographical map of the flake obtained at increasing temperature. All the flakes, grown at the different temperatures, present fractal saw toothed edges, which are probably mediated by a diffusion-limited-aggregation (DLA) regime[47,48] In addition, the flake edges present nanoparticles decoration. Such effect is directly connected with the employment of the liquid precursor, since the nanoparticles are composed of precursor byproducts, as sodium oxide (Na$_2$O).[49]

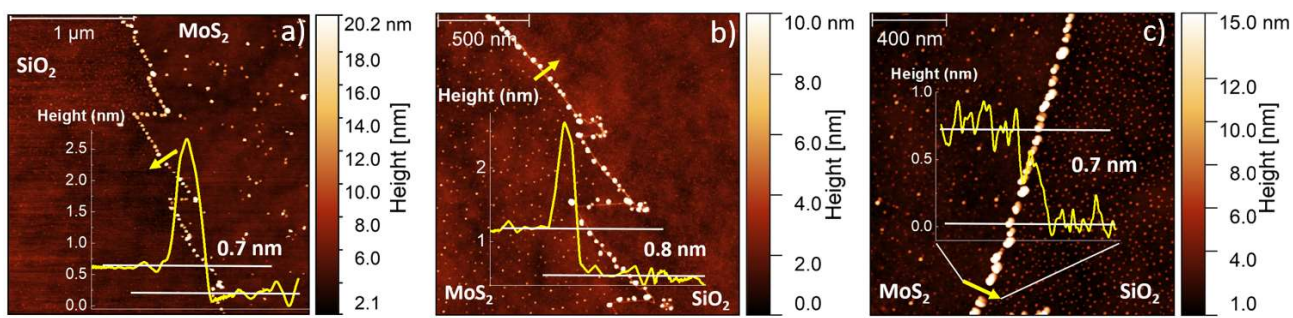

**Figure 2.** AFM topographic maps of the edges of the flakes grown at increasing temperatures: a) 810°C, b) 820°C and c) 830°C. The height profiles are obtained where the yellow line is reported on each map.

The AFM topographic maps, acquired on the edge of the flakes, are reported in Fig. 2. The main purpose of such analysis is the evaluation of the number of layers of the MoS$_2$ flake and the morphological analysis of the nanoparticles (NPs) decorating the flake edges. The topographical analysis reveals the presence of nanoparticles of smaller size inside the MoS$_2$ flake and on the SiO$_2$ substrate. All the MoS$_2$ flakes are monolayer: the height profiles reported on the different panels, reveal a flake thickness of 0.7 - 0.8 nm, the standard thickness of MoS$_2$ monolayer.[50–53] Cross-sectional TEM analysis is reported in Fig. S1, confirming the monolayer nature of MoS$_2$ obtained at 830°C. It is worth noting that all the height profiles are acquired in area of the flake not affected by the nanoparticles and the flakes synthesized at 810°C and 820°C present a few nanometer thick step on the edge, similarly to WS$_2$ flakes grown by CVD process using tungsten liquid precursor.[44] The nanoparticles decorating the flake edges present similar diameter, (26 ± 7) nm, while the height increases, increasing the growth temperature. In fact, the height of the nanoparticles varies from (8 ± 2) nm at 810°C to (18 ± 5) nm at 830°C, indicating that the temperature has an important role in the formation of precursor byproducts. The density of the nanoparticles increases in case of the growth carried out at 830°C with an average linear density of nanoparticles of 32 NPs/μm, while the linear density decreases to 22 NPs/μm for synthesis at lower temperatures.

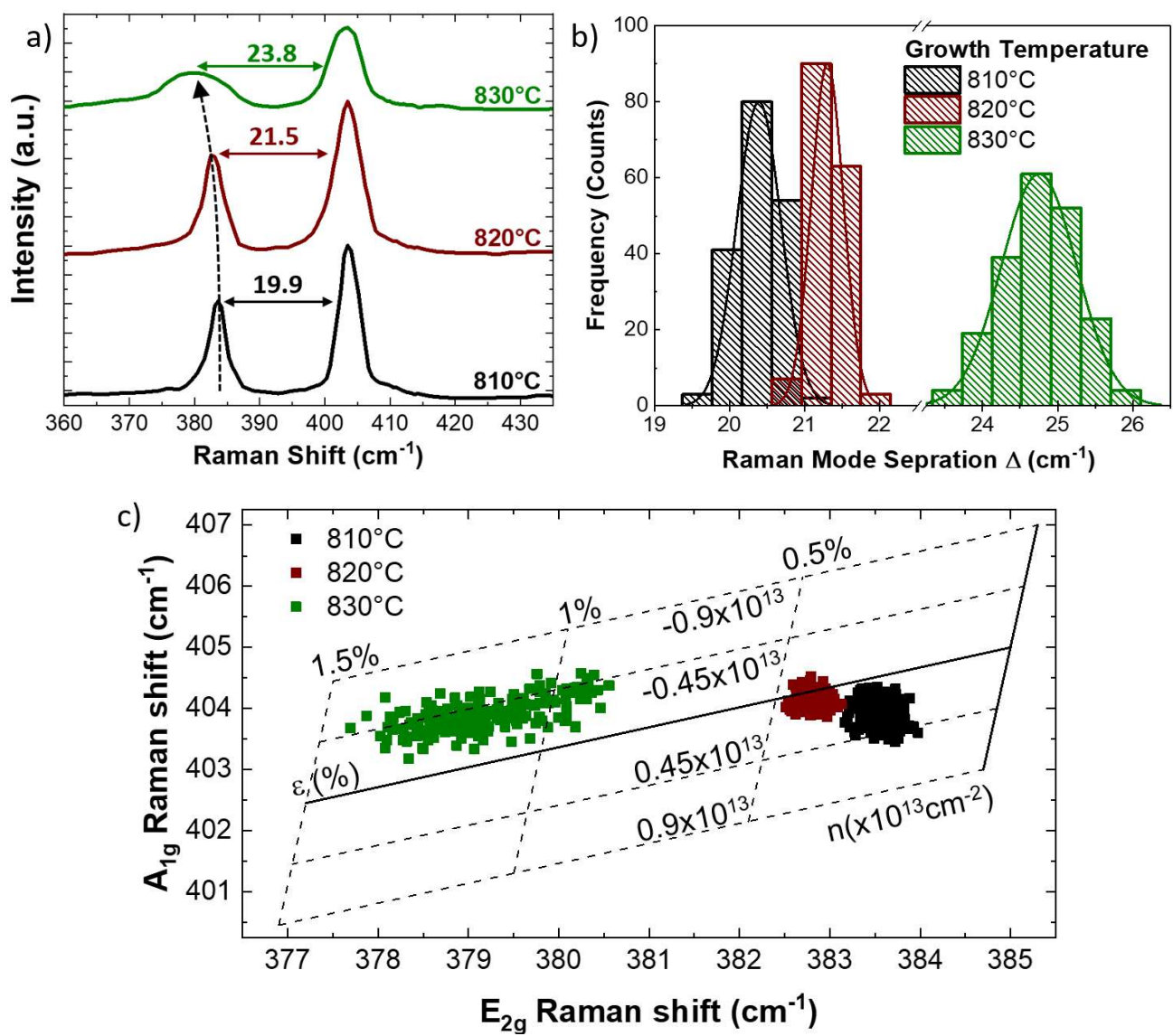

**Figure 3.** a) Representative Raman spectra of the MoS$_2$ monolayers obtained at increasing temperature: 810°C (black line), 820°C (red line), 830°C (green line). The spectra are vertically shifted for sake of clarity. The Raman mode separation is indicated for each spectrum. b) Histogram of the Raman mode separation for the different growth temperature. c) MoS$_2$ doping/strain correlation plot.

Figure3 presents the Raman characterization of the monolayer $MoS_2$, synthetized at increasing temperatures. Representative Raman spectrum (Fig. 3a), acquired at the center of the flake grown at 810°C (black line), presents the standard $MoS_2$ Raman modes, $E_{2g}$, at 383.6 cm$^{-1}$ and $A_{1g}$, at 403.5 cm$^{-1}$, with a Raman modes separation of 19.9 cm$^{-1}$. The $A_{1g}$ mode corresponds to the sulfur atoms oscillating in antiphase out-of-plane and the $E_{2g}$ mode is related to the sulfur and molybdenum atoms oscillating in antiphase parallel to the crystal plane.

The separation of the Raman modes below 20 cm$^{-1}$ is the standard benchmark of monolayer $MoS_2$.[23,54,55] However, this analysis of the $MoS_2$ monolayer grown at higher temperature demonstrates that both the $MoS_2$ Raman modes are shifted. In fact, in case of the flake grown at 820°C (red line) the $E_{2g}$ and the $A_{1g}$ modes appear at 382.5 cm$^{-1}$ and at 404 cm$^{-1}$, respectively. The separation of the Raman mode is 21.5 cm$^{-1}$, a value standardly reported for bilayer $MoS_2$.[54,55] For the flakes grown at the 830°C (green line) the $A_{1g}$ mode is still set at 404 cm$^{-1}$, while the $E_{2g}$ presents an additional shift down to 380.2 cm$^{-1}$. The separation of the Raman modes is 23.8 cm$^{-1}$, the standard benchmark of few-layers $MoS_2$.[54,55] The larger shift of the $E_{2g}$ mode is a symptom of the presence of strain because the $A_{1g}$ mode is less affected than the $E_{2g}$ mode, being the $A_{1g}$ mode related to the out-of-plane vibration with respect to the $E_{2g}$ mode. In particular, the different response to strain of the two $MoS_2$ Raman modes is reflected in the different Grüneisen parameter of each vibrational mode: $\gamma_{E2g}$= 0.68 and $\gamma_{A1g}$= 0.21.[26,39,56] Another clue of the presence of built-in strain applied to the $MoS_2$ grown at high temperature is the broadening of the $E_{2g}$ mode, whose full width half maximum (FWHM) from 2.5 cm$^{-1}$ (810°C grown monolayers) up to 10.4 cm$^{-1}$ (830°C grown monolayer)[26], in fact the application of strain induces the degeneracy breaking of the E Raman mode. The breaking of the degeneracy causes the splitting of the $E_{2g}$ mode in case of the presence of uniaxial strain larger than 0.8%.[21,27] Figure 3b reports the statistical analysis of the Raman modes separation, obtained by the Raman maps of a single flake shown in Supplementary Figure S2. The Gaussian distribution of the Raman modes separation is obtained by fitting the histogram data. In the case of the flakes synthetized at 810°C, the distribution is centered at 20.4 cm$^{-1}$ with a FWHM of 0.7 cm$^{-1}$ and it becomes even sharper in case of the synthesis at 820°C with a FWHM of 0.5 cm$^{-1}$, being centered at 21.3 cm$^{-1}$. The largest distribution of the Raman mode separation is a FWHM of 1.4 cm$^{-1}$ in case of the 830°C grown monolayers, where the Gaussian fitting reveals that the center is at 24.8 cm$^{-1}$. The statistical analysis of the Raman modes separation is necessary to clarify the discrepancy with the AFM analysis reported in Fig. 2, in fact all the flake analyzed are monolayer in nature whereas the Raman suggests a different number of layers for the increasing growth temperature. This is clear in regards of the different shift of the two Raman modes when strain is applied to $MoS_2$, because the $E_{2g}$ mode is more affected than the $A_{1g}$ mode, as previously explained. Therefore, we can assess that the presence of built-in strain in the MoS$_2$ synthesis with liquid precursors, makes unreliable the Raman method for the evaluation of layers' number, using the separation of the vibrational modes. In order to quantify the built-in strain and the modification of doping, we employ the MoS$_2$ correlation plot of the Raman shifts of the E$_{2g}$ and A$_{1g}$ modes, also known as the $\varepsilon$-$n$ system.[57] This method allows to disentangle and to quantify the strain and doping variations and it is normally employed for studying the growth induced strain[58], the effect of different growth substrates[45] or in MoS$_2$ based van der Waals heterostructures[59,60]. The full lines represent the zero strain and zero doping lines, while the dashed lines correspond to iso-strain and the iso-doping lines, calculated following the insights of previous works.[39,61] In addition, it is worth mentioning the importance of the origin of the $\varepsilon$-$n$ system: we set the zero strain and charge neutrality phonon frequencies 385 cm$^{-1}$ for the E$_{2g}$ mode and 405 cm$^{-1}$ for the A$_{1g}$ mode, evaluated in case of CVD grown MoS$_2$ suspended monolayer membrane.[39] The data regarding the MoS$_2$ monolayers grown at 810°C present a round distribution revealing an average value of (0.29 ± 0.06)% of tensile strain and an average positive charge concentration of (0.23 ± 0.22) x 10$^{13}$ cm$^{-2}$, while in case of the monolayer synthetized at 820°C, the data distribution is less dispersed and the average tensile strain increases up to (0.44 ± 0.12)% and the charge concentration decreases close to neutrality, (0.06 ± 0.10) x 10$^{13}$ cm$^{-2}$. The average strain reaches its maximum in case of the MoS$_2$ monolayer obtained at 830°C, (1.15 ± 0.5)%, albeit the data present a linear distribution along the iso-doping line relative to an electron charge concentration -0.45 x10$^{13}$ cm$^{-2}$. In order to clarify the data spreading in the A$_{1g}$ versus E$_{2g}$ correlation plot, in case of the monolayers obtained at 830°C, the spatial distribution of strain is shown in Fig. 4. The methodology for the development of the strain maps is reported in Fig. S5 with the strain maps of the flakes grown at 810°C and 820°C. The Raman spectra in the amorphous carbon range are reported in Fig. S3 showing the absence of any amorphous carbon related peaks, that can arise due to organic compounds (i.e. iodixanol) in the precursor solution. In addition, the data regarding the MoS$_2$ MLs transferred on a clean SiO$_2$/Si substrate are reported in Fig. S4.

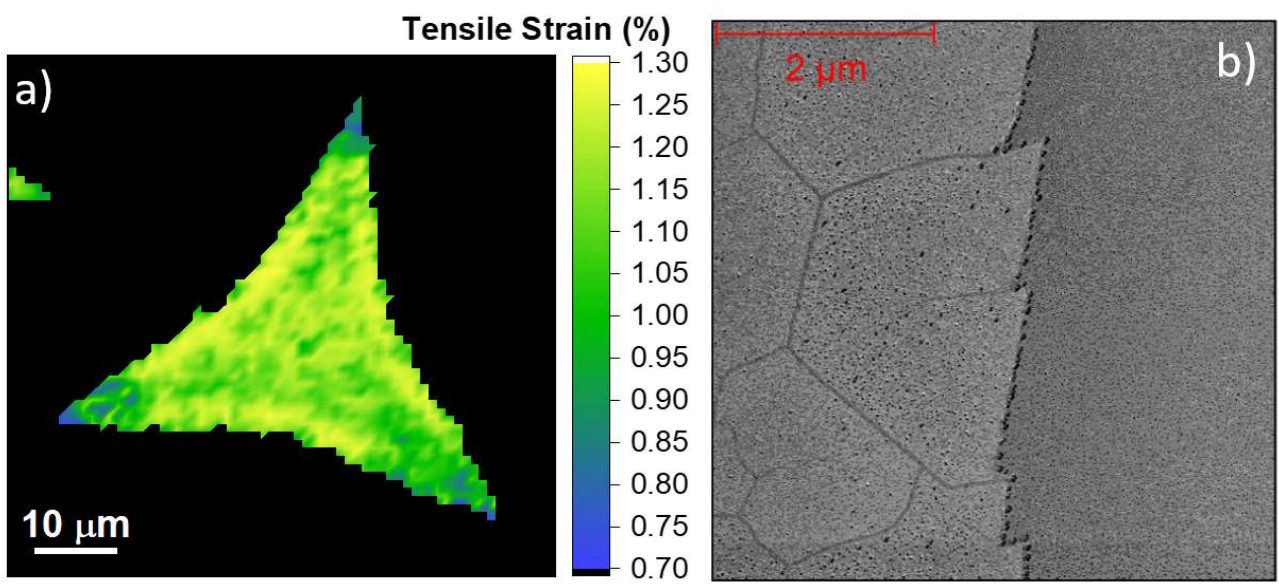

**Figure 4** Representative strain map of the MoS$_2$ monolayers obtained at 830°C (panel a) and the AFM phase map of the edge of the flakes (panel b).

The strain map of the 830°C grown monolayer (Fig. 4a) reveals that the strain is partially released in the triangular vertexes of the flake. This finding supports the spreading of the data reported in Fig. 3c. In fact, while in the central part of the flake is above 1%, the strain evaluation drops to 0.7% - 0.8%, close to the vertexes of the triangular monolayer. The strain release mechanism is based on the formation of nanocracks in such regions, demonstrated by AFM phase analysis (Fig 4b). The AFM phase analysis is employed to maximize the contrast of the nanocracks, the topographical map is reported in Fig S6 of the Supporting Information. A similar strain release mechanism has been previously reported in case of MoS$_2$ monolayer exposed to hydrogen or oxygen plasma.[62,63] The similarities rely mainly on the formation of six-fold hexagonal patterned net of nanocracks. The angle distribution of the lines relative to the edge of the MoS$_2$ crystal has preferential orientations of 0°, 60°, and 120° as shown in Fig. 4b, which indicates that the line patterns are associated with the crystallographic orientation of the MoS$_2$. This effect is important because it demonstrates that MoS$_2$ monolayer can sustain a limited amount of built-in tensile strain during the growth process, while the strain applied by external procedures can reach the 10%.[18]

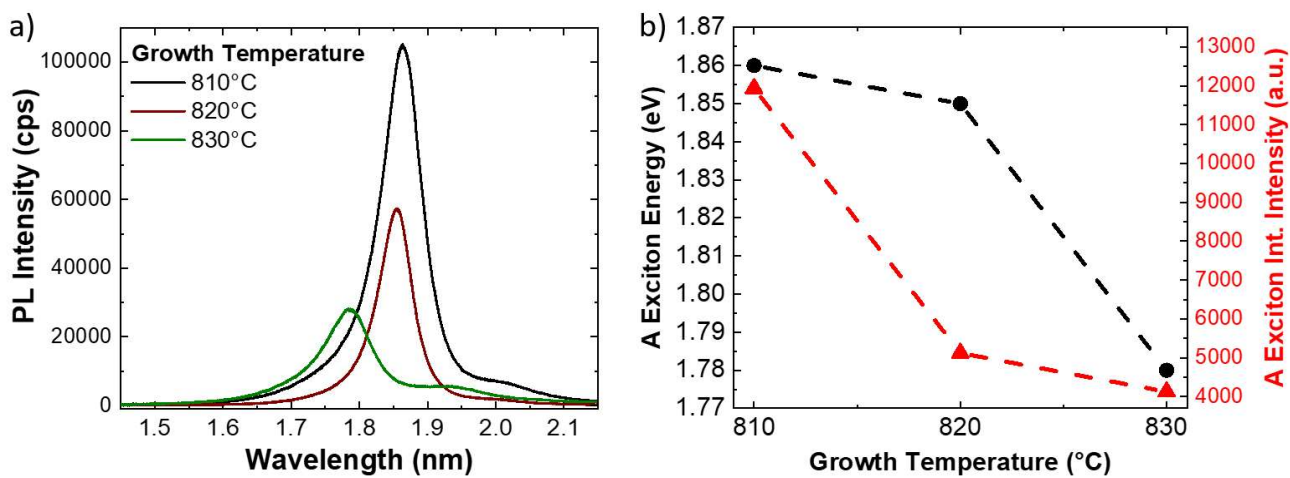

**Figure 5.** a) Representative PL spectra of the MoS$_2$ monolayers synthetized at 810°C (black line), 820°C (red line) and 830°C (green line), acquired at the center of the flake. b) Resuming behaviors of the A exciton emission energy (black dots) and PL intensity (red triangles) as function of the growth temperature. The dashed lines are guides for the reader eyes.

Another method for evaluating the strain in atomically thin semiconducting TMDs is the analysis of the photoluminescence (PL) emission. Several previous works have demonstrated the strain dependence of the PL emission energy and intensity.[21,22,27,39,42,64,65] In case of MoS$_2$ monolayer, the standard PL spectrum presents two main peaks attributed to the A and B excitons. The emission energy of the A exciton is reported to vary between 1.82 eV and 1.89 eV[21,27,65,66], while the B exciton energy varies between 1.97 eV and 2.05 eV[21,27,42,65,66]. In our particular case, Figure 5 presents the representative PL spectra of the MoS$_2$ monolayer grown at increasing temperatures, acquired at the center of the flake. The PL spectrum of 810°C grown ML shows one sharp peak at 1.86 eV and a faint shoulder in the high energy side, namely at 2.01 eV. Increasing the growth temperature at 820°C, the A exciton peaks suffers of 58% quenching of the integrated PL intensity and a slight red-shift of the emission energy (0.01 eV), being peaked at 1.85 eV. The red shift is even higher and the quenching more serious, in case of the 830°C grown monolayers, in fact the PL emission energy shifts down to 1.78 eV and quenching increases to 62%. It is worth noting that the quenching effect is limited by the broadening of the A excitonic PL peak. Fig. 5b resumes the PL A exciton emission energy and the PL integrated intensity as function of the growth temperature of the MoS$_2$ MLs. We consider a PL energy gauge factor of -99 ± 6 meV/%, measured in case of biaxial strain of CVD grown MoS2 monolayer[39], that is good agreement with theoretical prediction of 105 meV/%[67]. The biaxial strain, obtained by the PL shift, is 0.4% in case of the ML grown at 810°C, while it increases up to 0.5% for the 820°C synthetized MLs. The biaxial strain is maximized in case of the ML grown at 830°C with a value equal to 1.2%. It is worth noting that these values are calculated considering the peak energy of the A exciton for unstrained MoS$_2$ ML at 1.9 eV.[39,66]

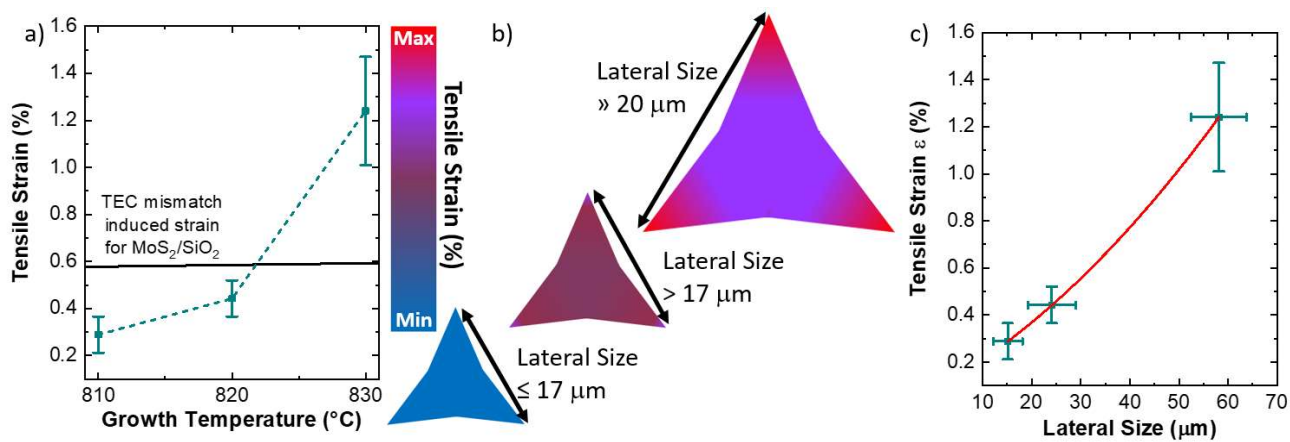

**Figure 6. a) Temperature dependent built-in strain values with a direct comparison with theoretical TCE mismatch strain of MoS$_2$/SiO$_2$ interface b) Sketch depicting the mechanism of the strain generation in the MoS2 ML with dependence on the flake size. c) Dependence of the strain on the lateral size of the flake, reporting the polynomial fit.**

Figure 6 proposes a mechanism for the large variation of the MoS$_2$ built-in strain with the reduced tuning of the growth temperature. This model is based on three main aspects: 1) the thermal expansion coefficient (TEC) mismatch between the two dimensional material and the substrate during the growth process,[4] 2) the employment of the liquid precursors in the CVD process,[49,58] 3) the strain dependence on the size of the MoS$_2$ ML synthetized at temperature beyond 800°C.[46,68] The direct comparison of the built-in strain values as function of the temperature with the theoretical TEC induced mismatch for the synthesis of MoS$_2$ on thermal silicon dioxide is shown in Fig. 6a. Thermal silicon dioxide presents a poor linear TEC equal to $\alpha_a = 0.24 \times 10^6$ K$^{-1}$ [69] while MoS$_2$ TEC is equal to $\alpha_a = 7.6 \times 10^6$ K$^{-1}$.[70] Based on this TEC mismatch, a degree of built-in tensile strain in the MoS$_2$ ranging from 0.577% at 810°C to 0.593% at 830°C, can be expected. Albeit, the experimental values show that the applied built-in strain are lower than the theoretical values of the TEC mismatch induced built-in strain in case of the synthesis at 810°C and 820°C. The partial release of strain can be due the presence of an interfacial layer of Na$_2$SiO$_3$, as demonstrated by the XPS analysis reported in the supporting information (Fig S8-S11), related to the use of liquid precursors. This demonstrates that the TEC mismatch has a limited effect on the built-in strain in MoS$_2$ flakes obtained with liquid precursors, while the main effect of the temperature is to increase the size of MoS$_2$ MLs. Therefore, the main cause of large built-in strain in MoS$_2$ ML grown at 830°C is related to the large lateral size of the sharp-vertex flakes. In fact, tensile strain-size dependence has been previously reported in case of $MoS_2$ ML grown at a temperature above 800°C, where the critical size limit for the appearance of strain is 17 µm.[46] In addition, the increase in size beyond the critical limit leads to spatial inhomogeneity of the built-in strain, where the edges[71,72] and the vertexes[46,73] are more affected. The possible mechanism is, therefore, resumed in Fig. 6b, where the MLs, obtained at 810°C presents a limited amount of homogeneous strain due to the size close to the critical limit of 17 µm. The $MoS_2$ MLs, presenting a size slightly greater than 17 µm, show an increasing strain in the body of the flake with an enhancement on the flake vertexes (See Fig. S5 for the strain map of 820°C grown flakes). The built-in strain is maximized with large inhomogeneity in the vertexes when the lateral size of the flake is much larger than 17 µm. The built-in strain value is more than two times the expected values for theoretical $MoS_2/SiO_2$ TEC mismatch induced strain, demonstrating that the size dependent strain is highly enhanced in case of large-area $MoS_2$ ML. It is worth noting that, in our particular case, the expected built-in strain in the triangular vertexes should be even higher despite the formation of nanocracks partially releases the accumulated strain. The dependence of the built strain as function of the ML size is reported in Fig. 6c. The polynomial fitting reveals that the semi-empirical law of the average built-in strain as function of the average size has parabolic behavior with the following parameters:

$$\varepsilon_{Built-in}= A \cdot d^2 + B \cdot d \qquad [1]$$

Where $\varepsilon_{Built-in}$ is the built-in tensile strain, $d$ is the lateral size of the ML, and $A$ and $B$ are the parameters obtained from the fitting procedures, imposing that the intercept is equal to zero and they are: $A$=7.3 x $10^{-5}$ %/µm$^2$, and $B$= 0.017 %/µm. Following this semi-empirical prediction, a $MoS_2$ ML with a lateral size of 100 µm should be affected by a 2.4% of strain in the central body of the flake. The built-in strain dependence on the ML size and its inhomogeneity can be attributed to the particular DLA growth regime, that gives rise to the fractal saw toothed edges. This dependence and inhomogeneity can be also enhanced by the particular growth temperature range (above 800°C) and the use of liquid precursors.

**Conclusion**

In conclusion, we demonstrate that the built-in strain of $MoS_2$ monolayers, grown on $SiO_2$/Si substrate using liquid precursors chemical vapor deposition, is mainly dominated by the size of the monolayer. Using correlative analyses of AFM, Raman and PL, we are able to highlight an inconsistency in the number of layers in $MoS_2$ flakes obtained by liquid precursors CVD. This leads to the identification of built-in strain dependent on the lateral size of $MoS_2$ monolayer. The built-in strain values are close to the prediction of thermal expansion coefficient mismatch for monolayer with a lateral size lower than 20 µm. The built-in strain is drastically increases for 60 µm sized monolayer, leading to 1.2% tensile strain with a partial release of strain close to the monolayer triangular vertexes due to formation of nanocracks. A semi-empirical model is defined for the possible prediction of the expected built-strain for a certain lateral size of MoS$_2$ monolayer without considering the possible release of strain due to the material cracking. The built-in strain dependence on the ML size and its inhomogeneity are possible attributed to the diffusion limited aggregation regime of the growth process, that gives rise to the fractal saw toothed edges of the MoS$_2$ monolayers.

**Experimental Section**

The chemicals composing the Mo liquid precursors are AMT (Sigma Aldrich purity 99.98%), NaOH (Carlo Erba) and Optiprep, a iodixanol based component provided by Serumwerk Bernburg AG normally employed in cell cultures.

MoS$_2$ flakes were synthesized at atmospheric pressure in an open tube using S powder with nitrogen as carrier gas. The employed substrates are commercial 300 nm thick SiO$_2$ coated highly conductive silicon wafer (Siltronic A.G.). Figure S12 shows a schematic illustration of the CVD reactor, i.e. an open tube with a diameter of about 1 inch heated in a two zones furnace. The sulfur boat is positioned in a low temperature zone (T = 180° C), while the growth substrate, after the spinning process of the Mo precursor solution, is placed in the high temperature zone (810°C < T < 830°C).[74,75]

AFM topography and phase maps were collected using a Bruker AFM operated in the Scan assist mode. Scanning Raman and photoluminescence spectroscopies were carried out with a Renishaw InVia system, equipped with confocal microscope, a 532 nm excitation laser and a 2400 line/mm grating (spectral resolution <1 cm$^{-1}$). All the analyses were performed with an 100X objective (NA=0.85), excitation laser power 500 µW, acquisition time 4s for each spectrum and pixel size of 1µm × 1µm.

Cross-sectional TEM analysis of a MoS$_2$ flake grown at 830 °C was performed on a TEM-lamella prepared by a Zeiss Auriga Compact Focused Ion Beam (FIB) system. In order to protect the atomically thin layer of MoS$_2$ from ion beam damaging[76], the specimen was coated with amorphous Carbon in a Balzers CED-010 setup prior to apply the standard procedure of FIB-lamella preparation. The lamella was then observed in a JEOL JEM 2200-FS microscope, operated at 200 kV.

We performed Xray Photoelectron Spectroscopy (XPS) in an ultra-high-vacuum (UHV) chamber using a VSW HA100 hemispherical electron energy analyzer with a PSP power supply and control.[77] We used a non-monochromatized Mgkα X-Ray source (photon at 1253.6 eV), with a final energy resolution at 0.86 eV. The Au4f 7/2 peak at 84.0 eV has been used as calibration for the binding energy (BE) scale. Core level lineshape analysis has been performed using Voigt functions with a Gaussian to Lorentian ratio of 30%, after the subtraction of a Shirley background. The typical precision for each component's energy position was ±0.05 eV. while for the area evaluation it was approximately ±2%.

**Acknowledgments.**

The authors wish to thank Prof. R. Fornari, (Department of Mathematical, Physical and Computer Sciences, University of Parma) for the useful discussion and support, Dr. G. Attolini for his contribution during the CVD growth of $MoS_2$ and Dr. G. Bertoni (CNR-Nano, Modena) for his assistance in the carbon coating process prior to the TEM lamella preparation.

# Supporting Information
# Synthesis of built-in highly strained monolayer $MoS_2$ using liquid precursor chemical vapor deposition

*L. Seravalli[1], F. Esposito[1,2], M. Bosi[1], L. Aversa[3], G. Trevisi[1], R. Verrucchi[3], L. Lazzarini[1], F. Rossi[1], F. Fabbri[4]*

[1] Institute of Materials for Electronics and Magnetism (IMEM-CNR), Parco Area delle Scienze 37/a, 43124 Parma, Italy

[2] Department of Mathematical, Physical and Computer Sciences, University of Parma, Parco Area delle Scienze 7/a, 43124 Parma, Italy

[3] Institute of Materials for Electronics and Magnetism (IMEM-CNR), FBK Trento unit, Via alla Cascata 56/C, 38123 Povo (Trento), Italy

[4] NEST, Istituto Nanoscienze – CNR, Scuola Normale Superiore, Piazza San Silvestro 12, 56127 Pisa, Italy

**List of the Figures:**

S1 Cross-sectional TEM analysis of 830°C grown $MoS_2$ ML

S2 Raman mapping of the $MoS_2$ modes separation and of the intensity of the $E_{2g}$ mode.

S3 Raman spectra in the range of the amorphous carbon.

S4 Transfer process and characterization of the $MoS_2$ MLs after transfer.

S5 Strain maps of the $MoS_2$ MLs grown at 810°C and 820°C.

S6 Topographical map of the $MoS_2$ ML affected by nanocracks.

S7-S8 XPS characterization of Mo precursor solution spunned on $SiO_2$/Si substrate undergone a high temperature treatment at 830°C.

S9 Schematic illustration of the CVD reactor.

**Figure S1: Cross-sectional TEM analysis of 830°C grown MoS2 ML.**

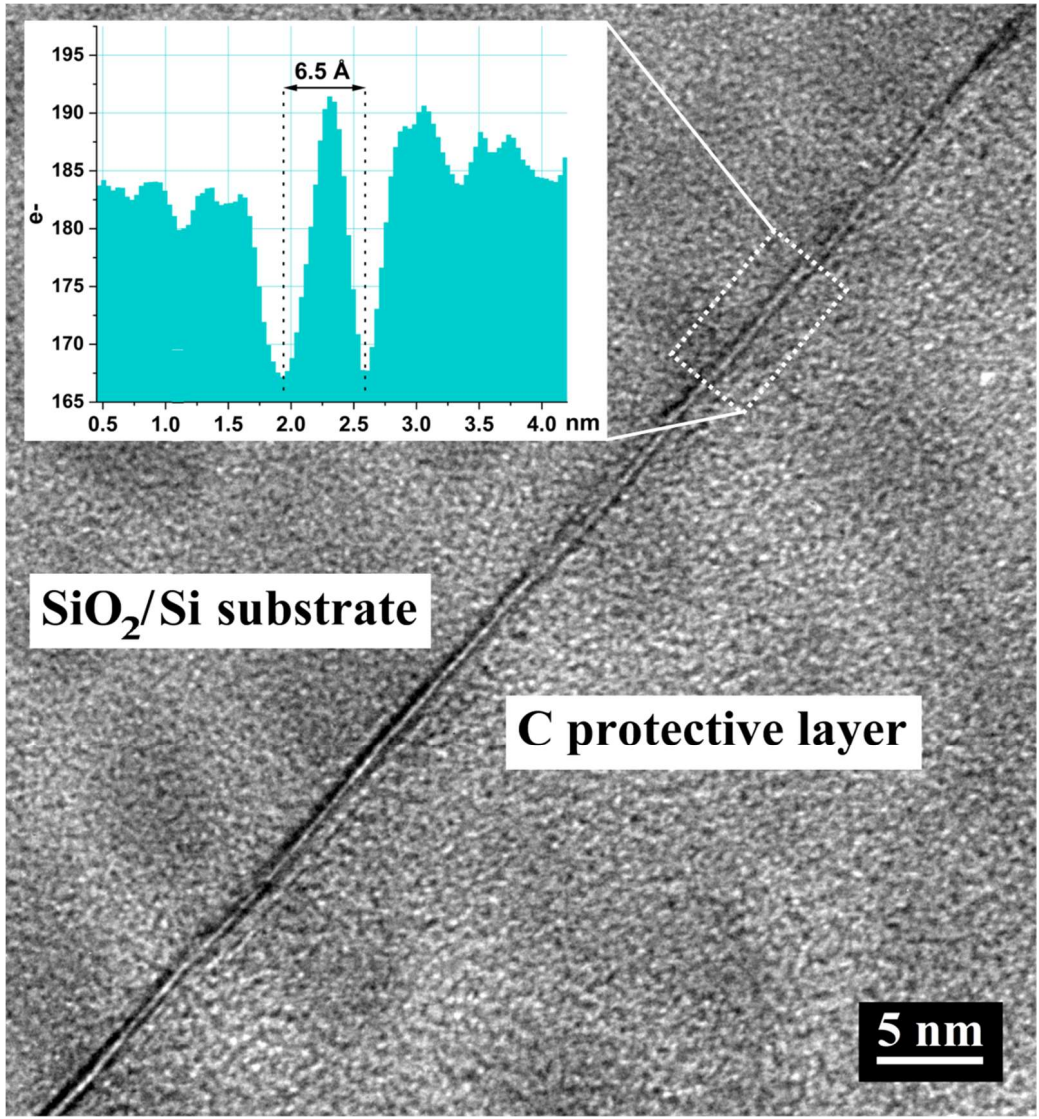

Fig. S1. A representative TEM image acquired at the interface, where MoS$_2$ is embedded between the amorphous SiO$_2$ and the protective Carbon coating. As shown from the intensity profile in the inset, a layer thickness of 0.65 nm is measured, that corresponds to the characteristic monolayer thickness of MoS$_2$.

**Figure S2: Raman mapping of the MoS$_2$ mode separation and of the intensity of the E$_{2g}$ mode.**

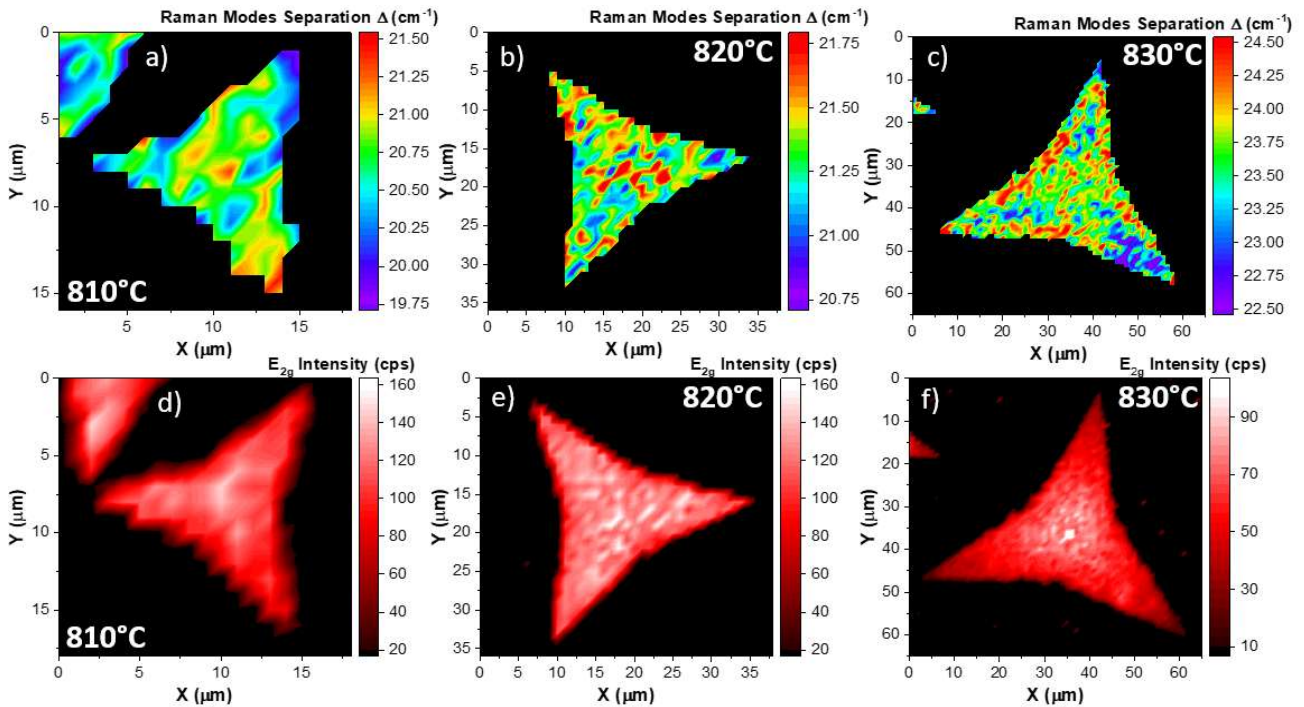

**Fig. S2 a), b) and c) Raman modes separation maps of the MoS$_2$ ML grown at 810°C, 820°C and 830°C, respectively. d), e) and f) E$_{2g}$ intensity maps of the MoS$_2$ ML grown at 810°C, 820°C and 830°C, respectively.**

The Raman mode separation maps reveal an enhanced mode separation, with increasing the growth temperature, as reported in Fig (S2a, S2b and S2c). The data reported here are employed for the statistical study shown in Fig. 3b. The intensity of the E$_{2g}$ mode are shown in Fig S2d, S2e and S2f for increasing growth temperature. The main effect is reported in Fig. S2f where the intensity of the E$_{2g}$ mode appears fainter in the vertexes of the triangular structure, an effect that can be related to the presence of defects, as the nanocracks reported in Fig. 4b.

**Figure S3: Raman spectra in the range of the amorphous carbon.**

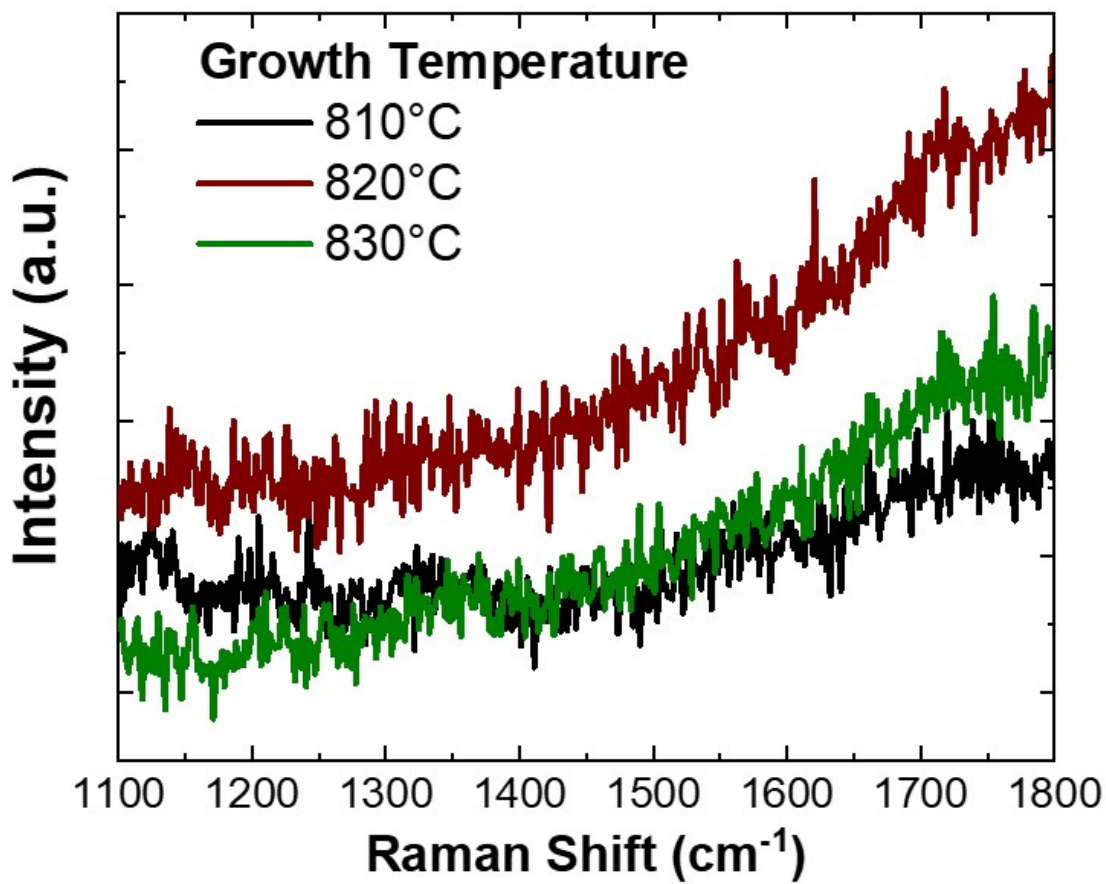

**Fig. S3 Raman spectra of the MoS$_2$ MLs grown at increasing temperature in the range of amorphous carbon.**

The Raman spectra in the range of amorphous carbon reveal no peaks that can be assigned to the presence of amorphous carbon in all the specimens in analysis. This result was previously reported for similar CVD process.[1]

**Figure S4: Transfer process and characterization of the MoS₂ MLs after transfer.**

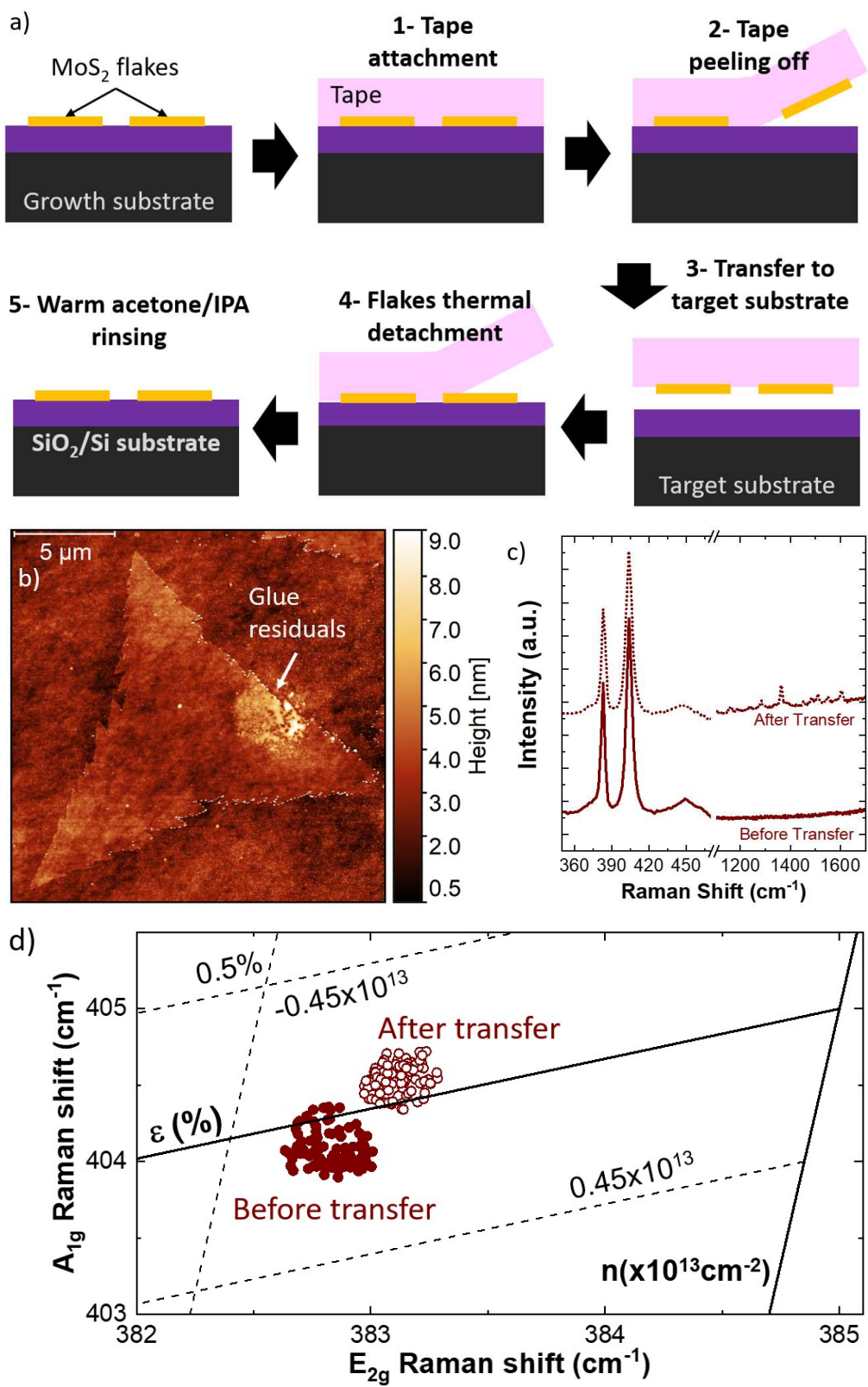

**Figure S4. a)** Graphical representation depicting the transfer process. **b)** AFM topological map of the MoS₂ ML, grown at 820°C, after the transfer. **c)** Representative Raman spectra before (full line) and after (dashed line) the transfer. **d)** MoS₂ doping/strain correlation plot, where the **full dots represent the data before the transfer while the open dots are acquired after the transfer.**

In order to clarify the effect of the substrate on the built-in strain, we transfer the $MoS_2$ MLs, grown at 820°C. It is worth mentioning that in the transfer of the 830°C synthetized MLs is hindered due to the presence of the nanocracks at the edge of the flakes. The transfer process is a modified dry viscoelastic stamping method[2–5], where the standard polydimethylsiloxane (PDMS) stamp is replaces with a polyethylene terephthalate (PET) tape with an acrylic adhesive. A similar approach has been widely employed in the transfer of graphene[6] and III-nitrides[7–9].

The transfer is based on different steps (Fig S4a). First, the PET tape is gently attached on the $SiO_2$/Si surface where the $MoS_2$ MLs are grown. Then the tape is peeled off very slowly and the MoS2 flakes remain attached to the tape. Then the tape is then attached on the acceptor surface, that in our case is a clean SiO2/Si substrate. The specimen with the tape on tops is then annealed at 125°C in order to melt the acrylic adhesive and the tape is then gently peeled off, releasing the $MoS_2$ MLs on top of the new substrate. Then the specimen is rinsed in warm acetone and IPA for the removal of the acrylic adhesive residuals. This particular novel transfer method is developed because the standard transfer approach, employing the poly(methyl methacrylate) (PMMA) spin coating on the specimen surface, is hindered by presence of the byproduct nanoparticles on the edge of the flakes. The AFM topological map, Fig S4b, reports the presence of some areas of the flake affected by thick dhesive residuals, nevertheless a large part of the nanoparticles decorating the flakes edge are absent. The comparison of the Raman spectra before and after the transfer reveals that the appearance of different Raman modes in the 1000-1600 cm$^{-1}$ range, related to the acrylic adhesive residuals (Fig. S4c).[10,11]

The analysis of the $MoS_2$ Raman modes (Fig.S4d), depicted in the $A_{1g}$ versus $E_{2g}$ Raman shift correlation plot, reveals that after the transfer process the strain is partially released in the MoS2 ML, decreasing down to 0.35% and with a concurrent neutralization of the free carrier concentration. This second effect can be related to the presence of the adhesive residuals on the $MoS_2$ surface, as previously demonstrated in case of graphene.[12,13]

**Figure S5: Strain maps of the MoS₂ MLs grown at 810°C and 820°C.**

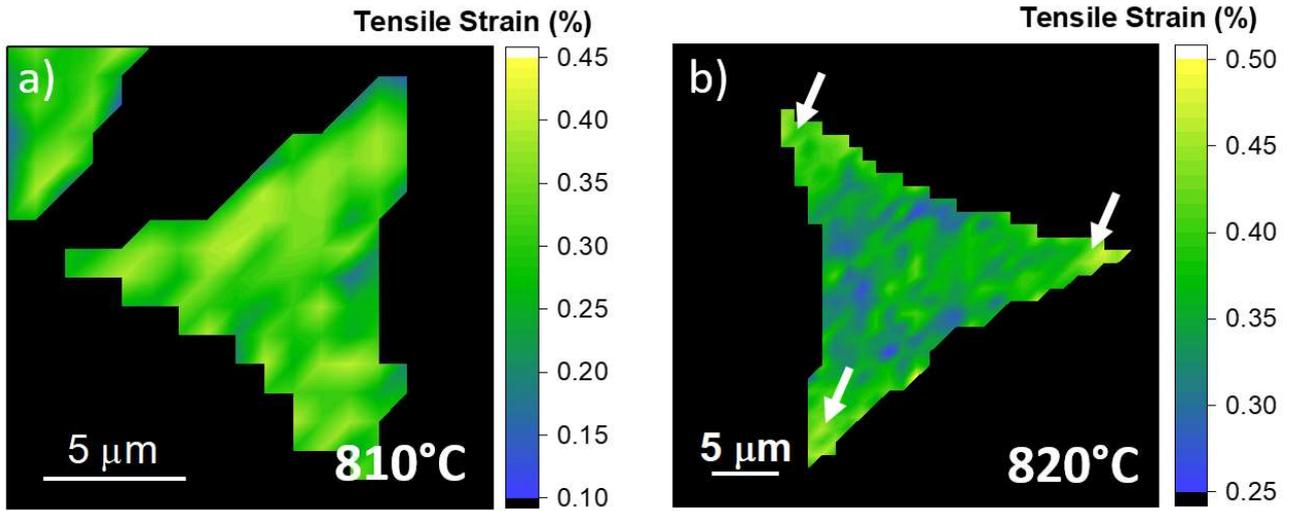

**Fig. S5 a) and b) Strain map of the MoS₂ ML synthetized at 810°C and 820°C with the same color code reported in Fig. 4 of the manuscript. The white arrows indicate the increase of strain at the vertexes of the vertex sharp triangular flake.**

The strain maps are obtained using the standard method proposed by A. Michail et al.[14,15]

$$\varepsilon = -\frac{\gamma(E')Pos(E')_0\boldsymbol{i} + \gamma(A_1')Pos(A_{1\prime})_0\boldsymbol{j}}{\sqrt{(\gamma(E')Pos(E')_0)^2 + (\gamma(A_1')Pos(A_1')_0)^2}}$$

The strain map (Fig S5a) of the MoS₂ ML grown at 810°C shows a tensile strain varying between 0.2% and 0.4%. In case of the MoS₂ ML grown at 820°C (Fig S5b), the strain map reveals that an increase of the tensile strain in the vertexes of the triangular monolayer, that reaches the maximum close to 0.5%, while in the body of the ML the tensile strain varies between 0.3% and 0.4%.

**Figure S6: Topographical map of the MoS$_2$ ML affected by nanocracks.**

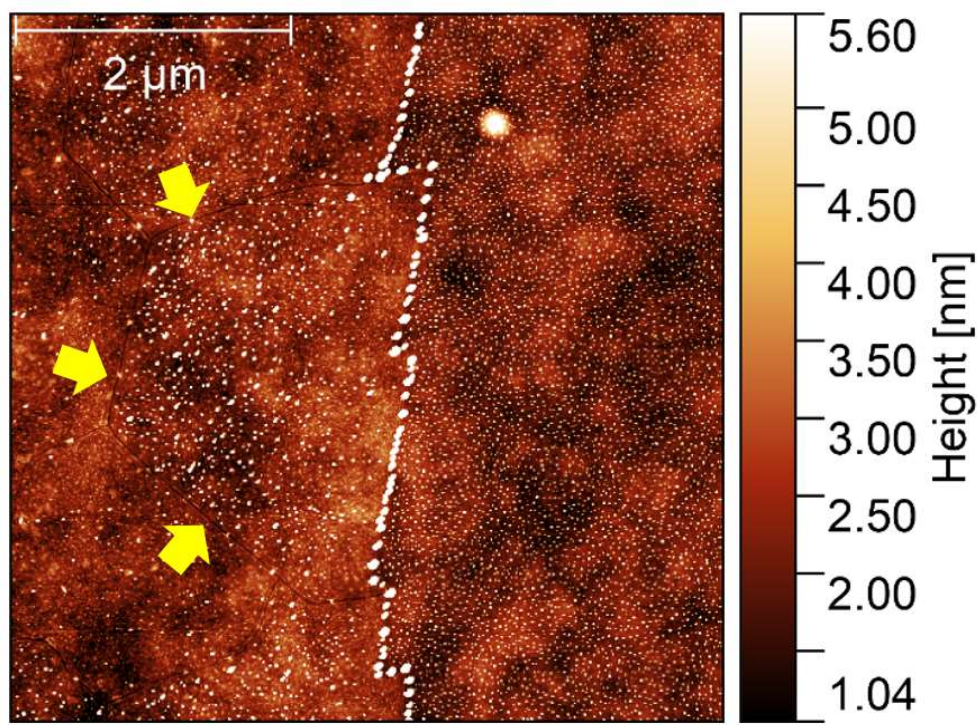

**Fig. S6 Topological map of the MOS2 ML area affected by nanocracks, indicated by yellow arrows.**

The detection of the nanocracks in AFM topographic mode is hindered by the presence of the byproduct nanoparticles inside the MoS2 monolayer. The yellow arrows highlight the nanocracks reported in Fig. 4b.

# Figure S7-S8: XPS characterization of Mo precursor solution spunned on a SiO$_2$/Si substrate undergone a high temperature treatment at 830°C

We performed X-Ray Photoelectron Spectroscopy (XPS) on the Mo precursor solution spunned on SiO$_2$/Si substrate after a high temperature treatment at 830°C to identify the surface chemical state of the compounds formed during such treatment. As a reference, we analyzed Na$_2$SiO$_3$ pure powder, purchased from Sigma-Aldrich, Merck Life Science S.r.l.. Despite the formation of Na$_2$MoO$_4$ is cited as possible reaction product between MoO3 and NaOH at high temperature[16], we do not detect any evidence of sodium molybdate from XPS. Na$_2$MoO$_4$ pure powder was analyzed for reference in the same UHV apparatus and compared to results from the treated surface ("Mo precursor solution" sample), but lineshape analysis is not consistent with this compound.

Long range spectrum of the treated substrate (Fig. S7, black curve) shows the presence of sodium (Na1s), oxygen (O1s), carbon (C1s), molybdenum (Mo3d), and silicon (Si2p). Mo3d core levels revealed the presence of two doublets (Mo3d 5/2, 3/2 due to spin orbit coupling) located at 232.2-235.3eV and 230.4-233.5eV, related to $MoO_3$ and $MoO_2$ molybdenum oxides (Fig. S8a)[17,18]. Si2p shows a main feature at 103.5 eV, due to $SiO_2$ and a minor component at 102.5 eV, that is consistent with the presence of $Na_2SiO_3$ (Figure S8b) and S8c), bottom curves). The Na/Si ratio is 2.1±0.05 and Na1s-Si2p energy difference is 969.4 eV (Na1s is located at 1072.1 eV), in good agreement with the results of sodium silicate analyzed as reference.

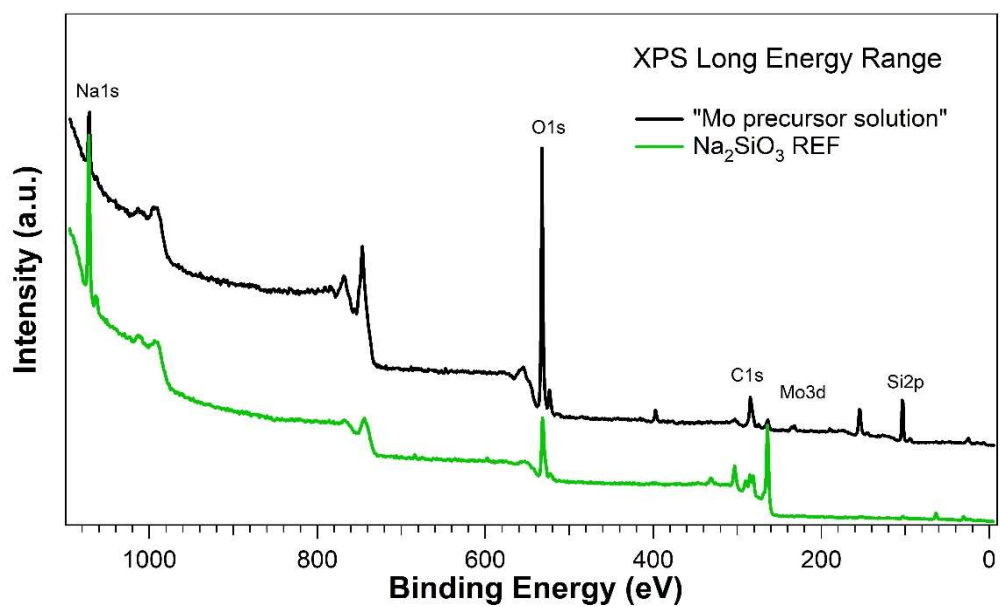

**Fig. S7: Long range XPS spectra of the Mo precursor solution on $SiO_2$ surface after a treatment at 830°C (black curve), and pure $Na_2SiO_3$ for reference (green curve).**

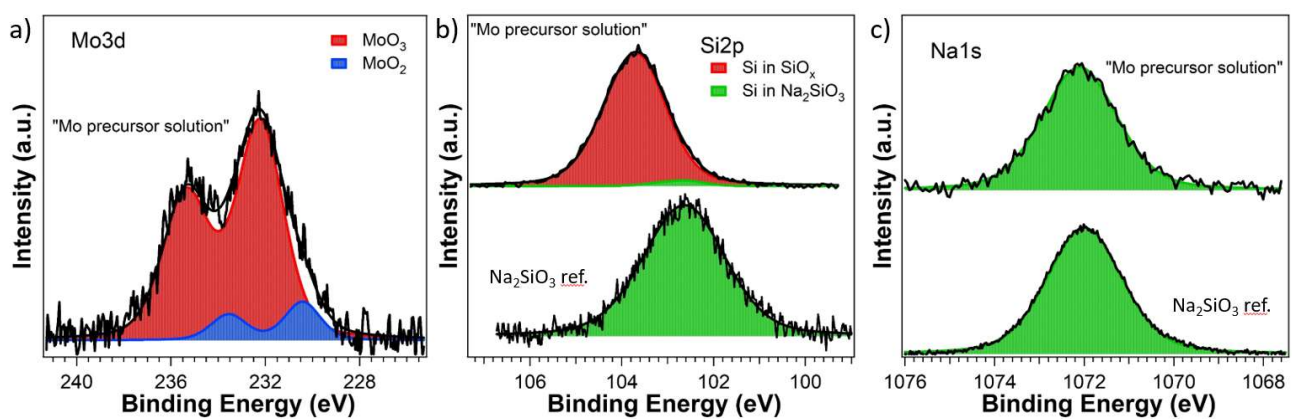

**Fig. S8 XPS core level spectra of Mo3d a), Na1s b), and Si2p c) of the Mo precursor solution on $SiO_2$ surface after a treatment at 830°C. In b) and c), the corresponding core levels of pure $Na_2SiO_3$ are shown for reference (bottom curves). Spectra are normalized in height.**

**Figure S9: Schematic illustration of the CVD reactor.**

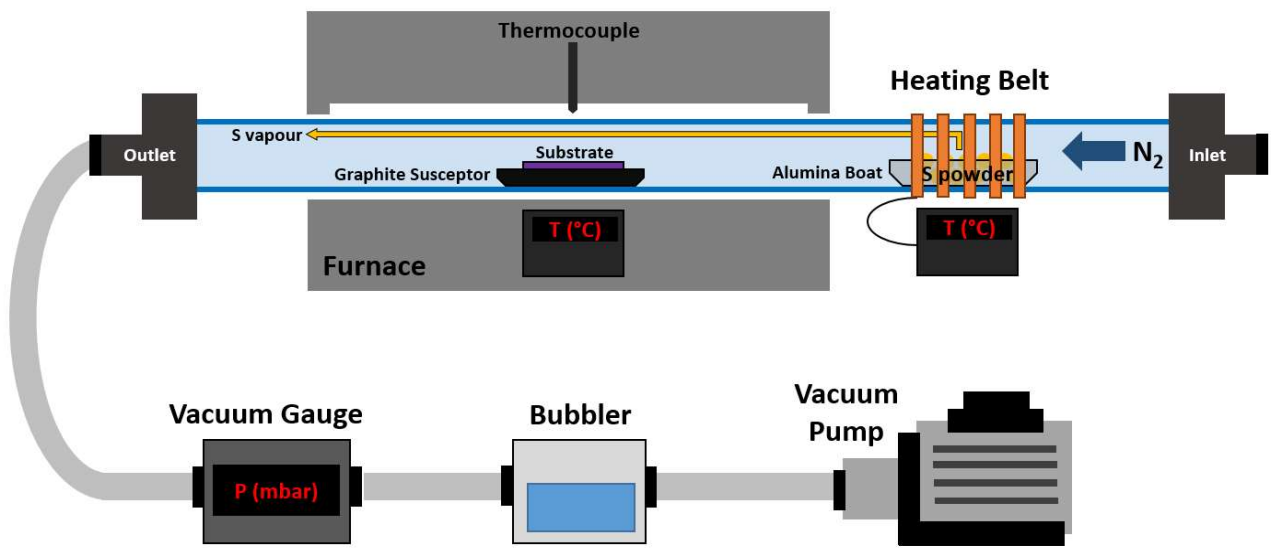

**Fig. S9 Schematic illustration of the CVD reactor, employed in this work.**

The schematic illustration of the CVD reactor is shown in Fig. S11. The low temperature zone is heated up using a heating belt to reach the desired temperature. The sulfur powder is placed on an alumina boat while the growth substrate, where the Mo precursor solution is spun, is positioned on a graphite susceptor for a uniform heating.